\documentclass[fleqn,12pt,twoside]{article}
\usepackage{espcrc1}

\usepackage{graphicx}
\usepackage[figuresright]{rotating}

\newcommand{\AmS}{{\protect\the\textfont2
  A\kern-.1667em\lower.5ex\hbox{M}\kern-.125emS}}
\newcommand{\beq}{\begin {equation}}
\newcommand{\eeq}{\end {equation}}
\title{A study of neutron-deuteron scattering in configuration space}
\author{V.M. Suslov\address[MCSD]{Department of Physics, North Carolina Central
University, 1801 Faytteville Str., Durham, North Carolina 27707 USA}
\thanks  {Also at Department of Mathematical and Computational Physics,
Sankt-Petersburg State University, 198504 Ul'yanovskaya Str.1, Petrodvorets
St.Petersburg, Russia. Electronic address: suslov@mph1.phys.spbu.ru}
        M.A. Braun\addressmark[MCSD] \address{Department of High Energy Physics,
Sankt-Petersburg State University, 198504 Ul'yanovskaya Str.1,
Petrodvorets St.Petersburg, Russia.},
        I.N. Filikhin\addressmark[MCSD]
        and
        B. Vlahovic\addressmark[MCSD]
\thanks {This work was supported by the Department of Defense and NASA through
Grants, W911NF-05-1-0502 and NAG3-804, respectively.}}
\begin{document}
\maketitle
\begin{abstract}
         A new computational method for solving the configuration-space
Faddeev equations for the breakup scattering problem \cite{Sus} has
been applied to  $nd$ scattering both below and above the two-body
threshold. To perform numerical calculations for a general nuclear
potential and with arbitrary number of partial waves retained, we
use the approach proposed in \cite{MGL}. Calculations of the
inelasticities and phase-shifts, as well as cross-sections and
analyzing powers for elastic $nd$ scattering at $E_{lab}$ =14.1 MeV
were made with the charge independent AV14 potential. The results
are compared with those of other authors and experimental data.
\end{abstract}

\section{Formalism}

 The Faddeev equations projected onto the MGL basis \cite{MGL} are :
\[
\Big[E+\frac{\hbar^2}{m}(\partial_x^2+\partial_y^2)-v_{\alpha}^{\lambda l}
(x,y)\Big]
\Phi^{\lambda_0,s_0,M_0}_{\alpha}(x,y)\]
\beq
\label{MGLeq}
=\sum_{\beta}v_{\alpha\beta}(x)\Big[\Phi^{\lambda_0,s_0,M_0}_{\beta}(x,y) +
 \int_{-1}^1du \sum_{\gamma} g_{\beta\gamma}(x,y,u)
 \Phi^{\lambda_0,s_0,M_0}_{\gamma}(x',y')\Big].
\eeq Here Greek subindexes denote state quantum numbers:
$\alpha=\{l,\sigma,J,s,\lambda,t\}$, where $l$, $\sigma$, $J$ and
$t$ are the orbital, spin, total angular momenta and isospin of a
pair of nucleons, $\lambda$ is the orbital momentum of the third
nucleon relative to the c.m. of a pair nucleons, and $s$ is the
total "spin"(${\bf s =1/2 + J}$). ${\bf M = \vec \lambda + s}$ is
the three-particle angular momentum. The geometrical function
$g_{\beta\gamma}(x,y,u)$ is the representative of the permutation
operator $P^++P^-$ in MGL basis (details see in \cite{MGL}). In Eq.
(1) $v^{\lambda l}_{\alpha}$ is the centrifugal potential, and
nucleon-nucleon potentials are $ v_{\alpha\alpha'}(x)
=\delta_{\lambda\lambda'} \delta_{ss'}\delta_{\sigma\sigma'}
\delta_{JJ'}v^{\sigma J}_{ll'}$, where $v^{\sigma J}_{ll'}$ are the
potential representatives in the two-body basis ${\cal
Y}^{JJ_z}_{l\sigma} ({\bf\hat{x}})$.
\newpage
     The asymptotic conditions for the radial parts are
\beq
\label{as1}
\Phi^{\lambda_0s_0M_0}_{\alpha}(x,y)=\delta_{\lambda\lambda_0}\delta_{ss_0}
\delta_{\sigma 1}\delta_{J1} \hat{j}_{\lambda}(qy)\psi_l(x) +
ia^{M_0}_{\lambda s\lambda_0 s_0} h_{\lambda}^{(+)}(q y)
\psi_l(x)+O(y^{-1})
\eeq
 and
\beq
\label{as2}
\Phi^{\lambda_0s_0M_0}_{\alpha}(x,y)=A^{M_0}_{\alpha,\lambda_0 s_0}
\frac{e^{i\sqrt{E}X}}{\sqrt{X}} +O(X^{-3/2}),\ \ X^2=x^2+y^2.
\eeq
In these formulae $\hat{j}_{\lambda}(qy)$ is the regularized spherical Bessel
function, $\psi_l(x)$ is the deuteron wave function ($l$=0,2),
$h_{\lambda}^{(+)}(q y)$ is the regularized Hankel function,
$a^{M_0}_{\lambda s\lambda_0 s_0}$
and $A^{M_0}_{\alpha,\lambda_0 s_0}$ are the elastic and breakup amplitudes,
respectively, $q$ an $E$ are the c.m momentum and final energy.
\vspace{-0.5cm}
\begin{table}[ht]
\caption {$n-d$ elastic amplitudes.}
\label{table:1}
\begin{tabular}{c||c|c||c|c}
\hline
$a^{M^{\pi}_0}_{\lambda s,\lambda_0 s_0}$  & Ref. \cite{PhysR} & present & Ref. \cite{PhysR} & present    \\
\hline
                         &    Elab=1 MeV   &                  & Elab=3 MeV     &                  \\
\hline
 $a^{1/2^+}_{1/20,1/20}$ &-0.291 + i 0.093 &-0.2903+ i 0.093  &-0.4691+ i 0.3272&-0.4679+ i 0.324  \\
 $a^{1/2^+}_{3/22,3/22}$ &-0.0175+ i 0.0003&-0.0173+ i 0.0003 &-0.0682+ i 0.0048&-0.0616+ i 0.004  \\
\hline
 $a^{1/2^-}_{1/21,1/21}$ &-0.0718+ i 0.0055&-0.0699+ i 0.0052 &-0.122 + i 0.0198&-0.120 + i 0.0191 \\
 $a^{1/2^-}_{3/21,3/21}$ & 0.2085+ i 0.0459& 0.208 + i 0.0457 & 0.3749+ i 0.176 & 0.3739+ i 0.1743 \\
\hline
 $a^{3/2^+}_{3/20,3/20}$ &-0.4985+ i 0.5383&-0.4985+ i 0.5382 &-0.3145+ i 0.888 &-0.3153+ i 0.8874 \\
 $a^{3/2^+}_{1/22,1/22}$ & 0.0101+ i 0.0001& 0.0108- i 0.0001 & 0.0420+ i 0.0018& 0.0437- i 0.0019 \\
 $a^{3/2^+}_{3/22,3/22}$ &-0.0187+ i 0.0004&-0.0181+ i 0.0004 &-0.0735+ i 0.006 &-0.0721+ i 0.0057 \\
\hline
 $a^{3/2^-}_{1/21,1/21}$ &-0.0718+ i 0.0052&-0.0716+ i 0.0052 &-0.1227+ i 0.016 &-0.1212+ i 0.0157 \\
 $a^{3/2^-}_{3/21,3/21}$ & 0.2392+ i 0.061 & 0.2401- i 0.0615 & 0.396 + i 0.1959& 0.3974- i 0.1978 \\
 $a^{3/2^-}_{3/23,3/23}$ & 0.0021+ i 7.1E-6& 0.0023+ i 7.6E-6 & 0.0158+ i 0.0003& 0.017 +i 0.0004  \\
\hline
\end{tabular}
\end{table}
\vspace{-0.5cm}
\section{Results}
In the present run we used the charge independent AV14 potential and included
all angular momenta of subsystems, $l$ and $\lambda$ $\leq$ 4,
total angular momentum of a pair nucleons $J \leq$ 3, and total three-body
angular momentum $M_0$ up to 9/2. The cutoff radius was taken 80 fm.
In Table~\ref{table:1}  our results on the $nd$ elastic amplitudes are given.
For comparison, accurate results of Bochum and Pisa group are also presented.

\begin{figure}[htb]
\vspace{-1.5cm}
\includegraphics[scale=0.55]{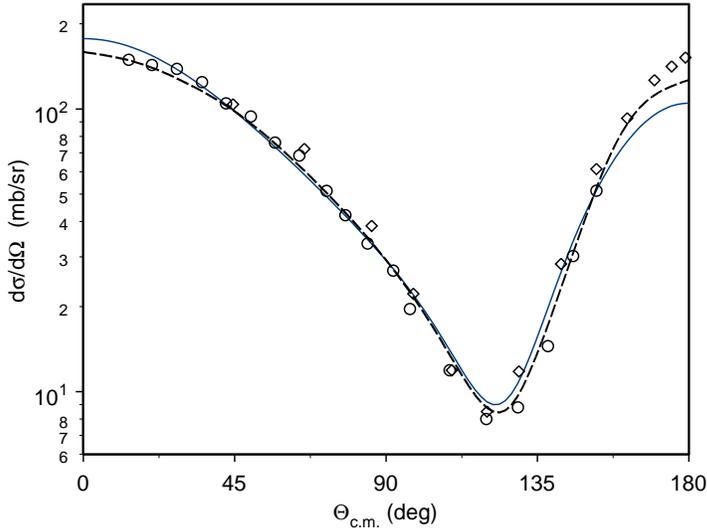}
\vspace{-1.2cm} \caption {\label{fig:dcs} $nd$ elastic differential
cross section at $E_{lab}$=14.1 MeV. The solid line is our results.
The dashed one corresponds to the Bochum group prediction
\cite{PhysR}. The data are from \cite{dif1} (open circles) and
\cite{b} (open diamonds).}
\end{figure}

\begin{figure}[htb]
\vspace{-2.75cm}
\begin{minipage}[t]{85mm}
\hspace*{-5mm} \includegraphics[scale=0.5]{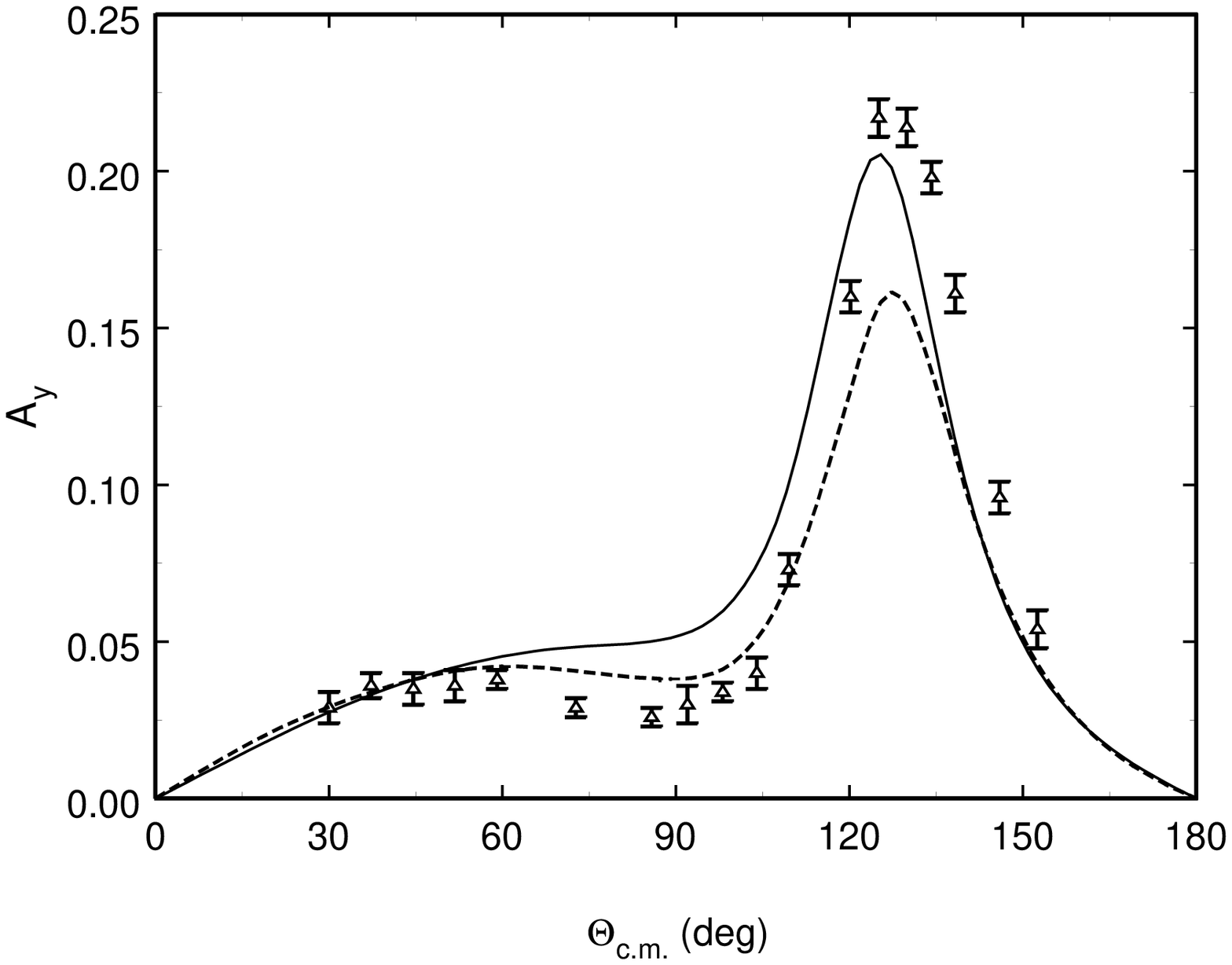}
\vspace{-2cm} \caption { \label{fig:Ay} The vector analyzing power
Ay at $ E_{lab}$ = 14.1 MeV. The solid line is our results. The
dashed one is the Bonn B prediction with $J\leq$3 of \cite{WitGl}.
The data are from \cite{FBS2}.}
\end{minipage}
\begin{minipage}[t]{85mm}
\includegraphics[height=82mm,width=82mm]{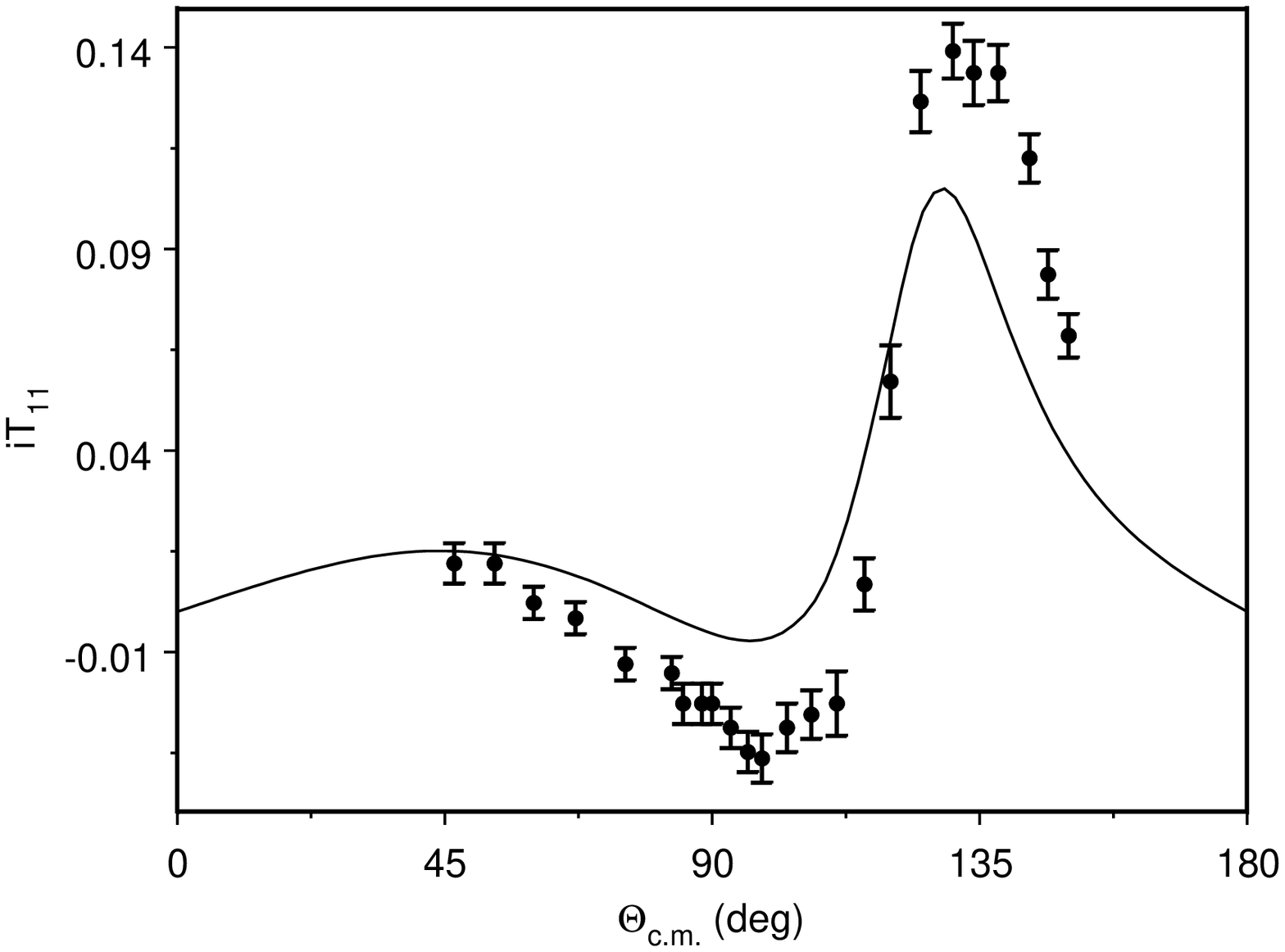}
\vspace{-1.cm} \caption {\label{fig:iT11} The deuteron analyzing
power $i\rm T_{11}$ at $E_{lab}$= 14.1 MeV. The solid line is our
results. The $pd$ data at $E_{lab}$=15.0 MeV are from \cite{PRL}.}
\end{minipage}
\end{figure}

 In Figure~\ref{fig:dcs} the elastic differential cross-section is shown
together with that of Bochum group \cite{PhysR} and the experimental data of
\cite{dif1,b}. Agreement is good except for angles in the vicinity
of 180 degrees. In Figures~\ref{fig:Ay} and \ref{fig:iT11} vector analyzing powers are shown
along with the experimental data.
\newpage
As one can see agreement with experimental data
is enough good in both cases. However one to take into account our limitations
on the maximal values for $M$, $l$ and $\lambda$. Convergence with these values
is presently under study.

   To check accuracy of our calculations we applied the optical
theorem. Presenting it in the form ${\rm Im}\,a=R.H.S$, where the
expression for $R.H.S$ may be found in ~\cite{MGL}, we get the
results shown in Table~\ref{table:2}.
\newpage
\begin{table}
\caption {The optical theorem results for $\rm E_{lab}$ = 14.1 MeV
and ${\bf M^{\pi}_0} = 1/2^+,3/2^+$}
\label{table:2}
\begin{tabular}{ccc|ccc}
\hline
 $ s_0, \lambda_0 $ & ${\rm Im} a^{M^{1/2^+}_0}_{\lambda_0 s_0,\lambda_0 s_0}$ &  R.H.S. &
 $ s_0, \lambda_0 $ & ${\rm Im} a^{M^{3/2^+}_0}_{\lambda_0 s_0,\lambda_0 s_0}$ &  R.H.S. \\
\hline
 \ \  1/2, 0 \ \ &  0.7172   &\ \  0.7182 \ \  & \ \ 3/2, 0 \ \ & 0.8942  &     0.8971\\
      3/2, 2     &  0.0258   &     0.0263      &     1/2, 2     & 0.0516  &     0.0522\\
        -        &     -     &       -         &     3/2, 2     & 0.0359  &     0.0363\\
\hline
\end{tabular}
\end{table}
Presenting the total cross-section $\sigma_{tot}$ as a sum of its elastic
$\sigma_{el}$ and inelastic  $\sigma_{in}$ parts we obtained values for them
shown in Table~\ref{table:3}.
\begin{table}[h]
\caption {Cross-sections of $nd$ breakup scattering for $\rm E_{lab}$ =14.1 MeV in [mb]}
\label{table:3}
\begin{tabular}{llll}
\hline
  $M_0 $ &$\sigma_{tot}$&$\sigma_{el}$ &$ \sigma_{in}$    \\
\hline
      $1/2$\ \ &  183.25   & 132.47      &   50.78    \\
      $3/2$   &   584.82    &   471.14    &   113.68   \\
      $5/2$   &    786.61  &    637.14   &     149.47\\
      $7/2$   &    804.25  &    648.20    &     156.05\\
\hline
              &  $824. \pm10$ & $694.\pm42$  &   $130.\pm43$ \cite{b}\\
              &  807. \cite{a}&              &   $172.\pm12$ \cite{c}\\
\hline
\end{tabular}
\end{table}
%
\section {Conclusion}

\noindent 1. The proposed method of the solution of the Faddeev
equations in configuration space by the direct integration of the
set of partial differential equations is fully feasible and
reliable, and give results of the same precision as obtained by
different methods.

\noindent 2. This shows that the method can be generalized to
include the Coulomb interaction in the $pd$ scattering below and
above the deuteron threshold, where others methods experience
considerable difficulties especially for the breakup channel.

\end{document}